\documentclass[prb,twocolumn,showpacs,preprintnumbers,amsmath,amssymb]{revtex4}

\usepackage{graphicx}
\usepackage{dcolumn}
\usepackage{bm}

\begin{document}

\preprint{APS/123-QED}

\title{Spin-modulated
 quasi-1D antiferromagnet LiCuVO$_4$}

\author{N. B\"{u}ttgen} \email{norbert.buettgen@physik.uni-augsburg.de}
\author{H.--A. Krug von Nidda}
\affiliation{Center for Electronic Correlations and Magnetism EKM,
Experimentalphysik V, Universit\"{a}t Augsburg, D--86135 Augsburg,
Germany}

\author{L.E. Svistov}\author{L.A. Prozorova}
\affiliation{P.L.Kapitza Institute for Physical Problems RAS, 119334
Moscow , Russia}

\author{A. Prokofiev}\altaffiliation[present address:]{Institut f\"{u}r Festk\"{o}rperphysik Technische Universit\"{a}t Wien, A--1040 Wien, Austria}
\author{W. A\ss{}mus} \affiliation{Physikalisches
Institut, Johann--Wolfgang--Goethe--Universit\"{a}t, D--60054
Frankfurt, Germany
}%

\date{\today}

\begin{abstract}
We report on magnetic resonance studies within the magnetically
ordered phase of the quasi--1D antiferromagnet LiCuVO$_4$. Our
studies reveal a spin reorientational transition at a magnetic field
$H_{c1}\thickapprox 25$~kOe applied within the crystallographical
$\mathbf{ab}$--plane in addition to the recently observed one at
$H_{c2}\thickapprox$75~kOe [ M.G. Banks et al., cond-mat/0608554
(2006)]. Spectra of the antiferromagnetic resonance (AFMR) along
low--frequency branches can be described in the frame of a
macroscopic theory of exchange--rigid planar magnetic structures.
These data allow to obtain the anisotropy of the exchange
interaction together with a constant of the uniaxial anisotropy.
Spectra of $^7$Li nuclear magnetic resonance (NMR) show that, within
the magnetically ordered phase of LiCuVO$_4$ in the low--field range
$H<H_{c1}$, a planar spiral spin structure is realized with the
spins lying in the $\mathbf{ab}$--plane in agreement with neutron
scattering studies of B.J. Gibson {\em et al.} [Physica B {\bf 350},
253 (2004)]. Based on NMR spectra simulations, the transition at
$H_{c1}$ can well be described as a spin--flop transition, where the
spin plane of the magnetically ordered structure rotates to be
perpendicular to the direction of the applied magnetic field. For
$H>H_{c2}\thickapprox$75~kOe, our NMR spectra simulations show that
the magnetically ordered structure exhibits a modulation of the spin
projections along the direction of the applied magnetic field $H$.
\end{abstract}

\maketitle

\section{INTRODUCTION}

\begin{figure}
\includegraphics[width=65mm]{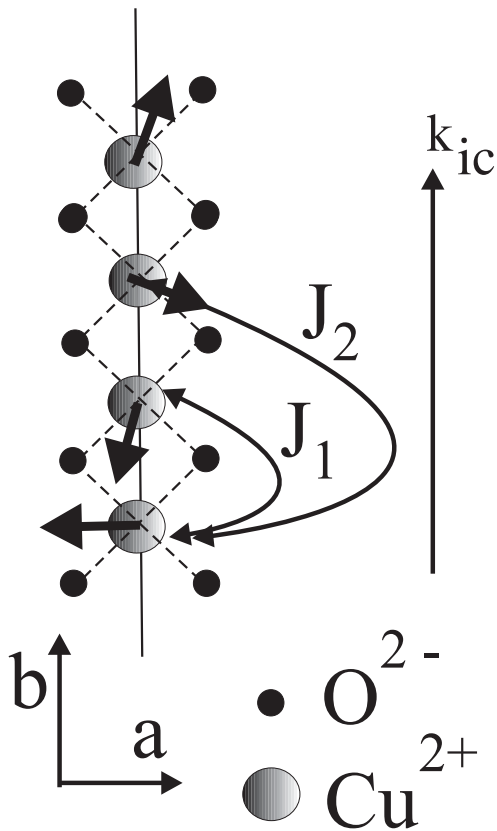}
\caption{Scheme of the spiral magnetic structure of LiCuVO$_4$(Ref.
\onlinecite{Gibs}. A fragment of one single chain of edge--sharing
CuO$_4$ units along with the dominant intrachain exchange integrals
$J_1$ and $J_2$ is shown.}
 \label{Fig_1}
\end{figure}

The quasi-one dimensionality (1D) of the spin system in LiCuVO$_4$
is provided by the magnetic ions of Cu$^{2+}$ (3$d^9$ configuration,
$S$=1/2). Lithium and copper ions share the octahedral sites in the
orthorhombically distorted spinel structure in that way that the
copper ions are arranged along chains separated by the nonmagnetic
ions of lithium, vanadium and oxygen. The temperature dependence of
the magnetic susceptibility exhibits a broad maximum at around
$T$=28 K, typical for low--dimensional antiferromagnets, and a sharp
anomaly at $T$=2.3 K, which is associated with the establishing of
three dimensional (3D) magnetic order. From the values of these
temperatures, the authors of Ref. \onlinecite{Vasi} evaluated the
intrachain and interchain exchange integrals of 22.5 K and 1 K,
respectively. Elastic neutron diffraction experiments \cite{Gibs} on
single crystals of LiCuVO$_4$ revealed that the 3D ordered phase
exhibits an incommensurate, noncollinear magnetic structure. The
scheme of this magnetic structure suggested is shown in figure
\ref{Fig_1}. According to that work,\cite{Gibs} the wave vector of
this magnetic structure is directed along the copper chains
($\mathbf{k}_{ic} \parallel \mathbf{b}$) and the value of the
ordered magnetic moments of the Cu$^{2+}$ ions amounts to
0.31$\mu_B$, lying within the $\mathbf{ab}$--planes. The
investigation of magnetic excitations in LiCuVO$_4$, using inelastic
neutron diffraction experiments,\cite{Ende} confirmed the quasi--1D
nature of this compound and brought to conclusion that the magnetic
incommensurate structure is due to an exotic ratio of intrachain
exchange integrals: the nearest--neighbor exchange integral
($J_1$=-18 K) is ferromagnetic and it is strongly dominated by the
value of the next--nearest exchange integral which is
antiferromagnetic ($J_2$=49 K, cf. Fig. \ref{Fig_1}). This unusual
hierarchy of exchange integrals is realized in LiCuVO$_4$ due to
indirect exchange interaction through the oxygen neighborhood and it
can be explained from the Goodenough--Kanamori--Anderson
rules.\cite{Mizu} Results of electron--band structure calculations
give the values of the exchange integrals $J_1$ and $J_2$ in a good
agreement with the values obtained experimentally.\cite{Ende}
The magnetic properties of such exchange--bonded chains of
Heisenberg spins with two exchange integrals $J_1$ and $J_2$ are
similar to the so called "zigzag"\ model where pairs of chains, with
an intrachain exchange integral $J_1$, are coupled via an interchain
exchange integral $J_2$. Such model was studied theoretically in the
limit of large spin values, as well as for $S$=1/2 in references
\onlinecite{Whit} and \onlinecite{Burs}. In these works it was shown
that such a ratio of the exchange integrals, which are obtained for
LiCuVO$_4$, results in an incommensurate magnetic structure. The
value of the incommensurate vector $\mathbf{k}_{ic}$ experimentally
obtained\cite{Gibs} is found to be between theoretical values in the
limit of large spin and for $S=1/2$ (cf. Ref. \onlinecite{Burs}).

Thus, LiCuVO$_4$ presents an example of a quasi--1D frustrated
antiferromagnet of spins $S$=1/2 where strong exchange interactions
yield an incommensurate noncollinear magnetically ordered structure
for temperatures $T<2.3$~K. In the present work, antiferromagnetic
resonance (AFMR) and nuclear magnetic resonance (NMR) on the nuclei
of nonmagnetic ions ($^7$Li and $^{51}$V) were studied for
temperatures within this magnetically ordered phase. Additionally,
the AFMR experiment is extended towards elevated temperatures
$T<30$~K where 3D order already is absent, but spin correlations
along copper chains are well developed. These temperatures are
within the range between the maxima of the temperature dependent
magnetic susceptibility $\chi(T)$ (cf. Fig. \ref{Fig_4}).

\section{SAMPLE PREPARATION AND EXPERIMENTAL DETAILS}

\begin{figure}
\includegraphics[width=65mm]{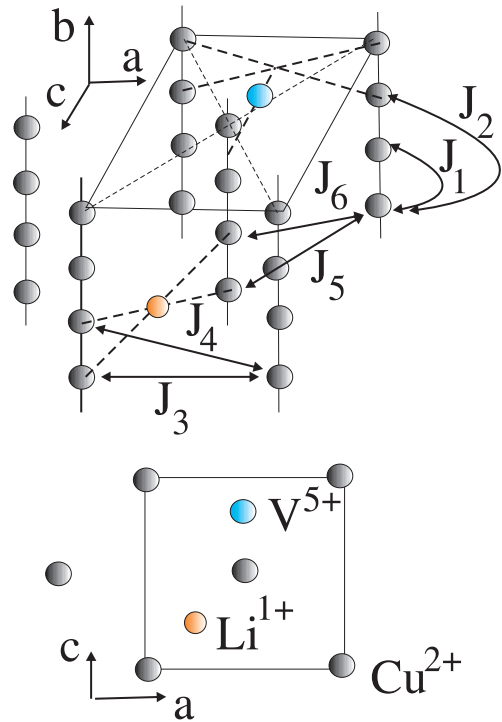}
\caption{Schema of the Cu$^{2+}$ sites in the crystal structure of
LiCuVO$_4$. The sites of Cu$^{2+}$ ions are marked by grey circles.
One of the lithium ions is marked by an orange circle and one of the
vanadium ions by a blue circle, respectively. Upper panel:
$J_1,\dots,J_6$ denote the main exchange integrals defining the
magnetic structure of LiCuVO$_4$ (Ref. \onlinecite{Ende}). Lower
panel: projection of the crystallographic structure on the
$\mathbf{ac}$--plane.}
 \label{Fig_2}
\end{figure}

LiCuVO$_4$ crystallizes in an inverse spinel structure AB$_2$O$_4$
with an orthorhombic distortion stipulated by a cooperative
Jahn--Teller effect, induced by Cu$^{2+}$ ions. The crystal
structure of LiCuVO$_4$ is described by space group Imma (Refs.
\onlinecite{Prok,Kegl1}). In Fig. \ref{Fig_2}, a schema of the
crystal structure of LiCuVO$_4$ is shown. The Cu$^{2+}$ ions, which
compose strongly exchange interacting magnetic  chains, are marked
with grey circles. For clearness these copper chains along the
crystallographic $\mathbf{b}$--direction are joined with a solid
line to guide the eye. Li$^{1+}$ ions occupy octahedrally
coordinated crystallographic equivalent sites and one of the lithium
ions is marked by an orange circle. V$^{5+}$ ions occupy the
tetrahedrally coordinated crystallographic equivalent sites and one
of the vanadium ions is denoted by a blue circle. The lattice
parameters in $\mathbf{a}$-- and $\mathbf{c}$--directions coincide
with the distance between nearest copper neighbors, but along the
$\mathbf{b}$--direction the lattice parameter equals twice the
distance between nearest copper neighbors.

\begin{figure}
\includegraphics[width=70mm]{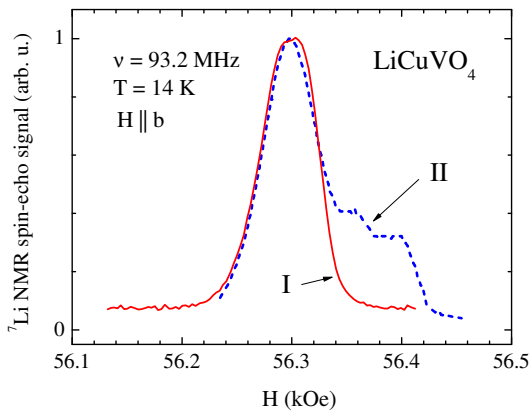}
\caption{$^7$Li NMR spectra of single crystals from two different
batches. All experiments of this work are performed with the single
crystal (I).}
 \label{Fig_3}
\end{figure}

Single crystals with the volume of some cubic millimeters were grown
as described in detail in reference \onlinecite{Prok}. As the ionic
radii of lithium and copper are very similar, it required special
efforts to satisfy the stoichiometry of this substance. Besides of
chemical methods described in Ref. \onlinecite{Prok2}, lithium
unsoundness was controlled additionally with the NMR study of $^7$Li
nuclei. Detailed NMR experiments in single crystalline LiCuVO$_4$ in
the paramagnetic phase are presented in reference \onlinecite{Kegl}.
Figure \ref{Fig_3} shows the NMR spectra obtained from two single
crystals with different quality which are denoted (I) and (II). The
spectrum of the single crystal (II) consists of several lines which
are due to defects as it is described in reference
\onlinecite{Kegl}. In the present work, all experiments are
performed with the single crystal (I) which exhibits a single $^7$Li
spectral line, demonstrating a unique lithium site. This single
crystal (I) was selected from the same batch as the single crystals
used in references \onlinecite{Gibs} and \onlinecite{Ende}.
The magnetic susceptibility was measured in the temperature range
$1.7 < T < 400$~K with a SQUID-magnetometer MPMS7 of Quantum Design.
The AFMR experiments were performed with a transmission--type
spectrometer using resonators in the frequency range
$18<\nu<110$~GHz. At elevated frequencies $65<\nu<250$~GHz, we
employed the wave--guide--line technique. The magnetic field of a
superconducting solenoid was varied in the range $0<H<90$~kOe.
Temperatures were varied within the range $1.2<T<30$~K. The NMR
experiments were performed with a phase coherent, homemade
spectrometer at radio frequencies within the range
$14.5<\nu<170$~MHz. We investigated the $^{7}$Li ($I$=3/2,
$\gamma$/2$\pi$=16.5466 MHz/T) and $^{51}$V ($I$=7/2,
$\gamma$/2$\pi$=11.2133 MHz/T) nuclei using spin--echo technique
with a pulse sequence 5$\mu$s--$\tau_D$--10$\mu$s, where the time
between pulses $\tau_D$ was 40 $\mu$s. The spectra were collected by
sweeping the applied magnetic field within $8<H<93$~kOe at constant
frequencies. Low temperatures $0.6<T<3.3$~K were achieved with a
$^3$He/$^4$He dilution refrigerator of Oxford Instruments. The
temperatures were stabilized with a precision better then 0.02~K.

\section{Magnetic susceptibility}
\begin{figure}
\includegraphics[width=75mm]{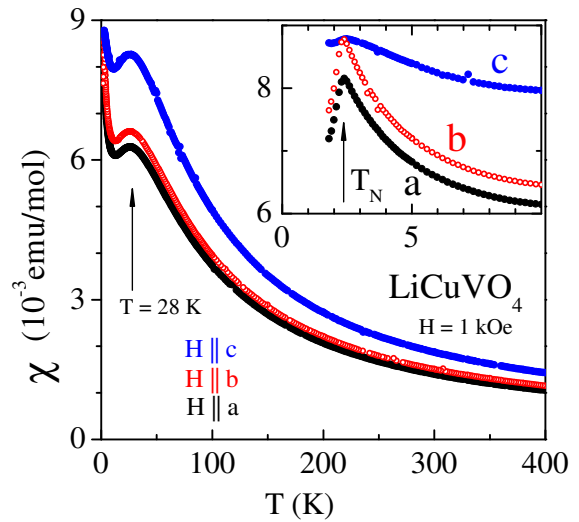}
\caption{Temperature dependences of the molar susceptibility
$\chi(T)$ of LiCuVO$_4$. The applied magnetic field of $H=1$~kOe was
directed along the principal axes $\mathbf{a,b}$ and $\mathbf{c}$ of
the single crystal. Inset: Magnification of the low--temperature
range for $T< 10$~K.}
 \label{Fig_4}
\end{figure}
Figure \ref{Fig_4} displays the temperature dependence of the molar
susceptibility $\chi(T)$ of LiCuVO$_4$ with the applied magnetic
field $H$ directed along the principal axes of the single crystal.
The molar susceptibility $\chi(T)$ exhibits two characteristic
maxima: the high--temperature maximum around $T\approx 28$~K is
usual for low--dimensional magnetic systems and it is associated
with the appearance of magnetic correlations within the copper
chains. The sharp low--temperature maximum (cf. the inset of Fig.
\ref{Fig_4}) is associated with the transition into a 3D
magnetically ordered phase at $T_{\rm N}=2.3$~K (Refs.
\onlinecite{Kegl,Krug}). The behavior of $\chi(T)$ with respect to
$\mathbf{H}\parallel \mathbf{a}$ and $\mathbf{H}\parallel
\mathbf{b}$ is almost the same, but it essentially differs from
$\chi(T)$ for $\mathbf{H}\parallel \mathbf{c}$. In the paramagnetic
phase and in the 3D ordered phase, the values of $\chi(T)$ for
$\mathbf{H}\parallel \mathbf{c}$ are remarkably enhanced compared to
$\chi(T)$ for $\mathbf{H}$ lying in the $\mathbf{ab}$--plane due to
the anisotropy of the $g$--tensor \cite{Krug}.

\section{AFMR}

\begin{figure}
\includegraphics[width=70mm]{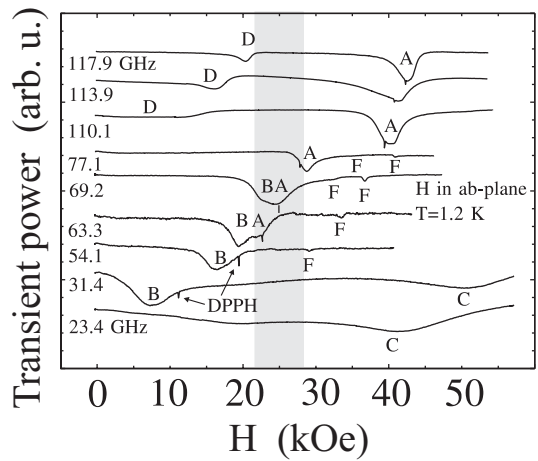}
\caption{Transient microwave power vs.\ applied magnetic field $H$
($\mathbf{H}\parallel \mathbf{ab}$--plane at 1.2~K). The signal of
the reference compound diphenyl--picryl--hydrazyl (DPPH) is marked.
For the denotations with capital letters A, $\dots$, F see text.}
 \label{Fig_5}
\end{figure}

\begin{figure}
\includegraphics[width=70mm]{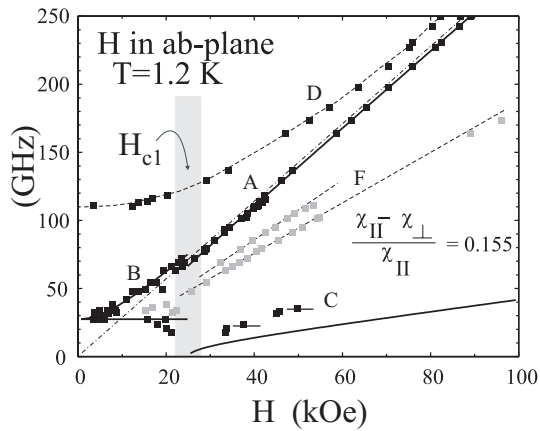}
\caption{AFMR spectra in the 3D ordered phase vs.\ applied magnetic
field $H$ ($\mathbf{H}\parallel \mathbf{ab}$--plane at 1.2~K). The
capital letters mark the different branches of the spectra according
to the denotations in Fig. \ref{Fig_5}.} \label{Fig_6}
\end{figure}

In Fig. \ref{Fig_5} we show examples of traces of the transient
microwave power which are obtained by sweeping the applied magnetic
field $H$. The orientation of $\mathbf{H}$ was chosen to lie within
the $\mathbf{ab}$--plane. All traces were recorded at the
temperature $T=1.2$~K, which is nearly two times less than $T_{\rm
N}$. The capital letters 'A', $\dots$, and 'F' denote absorption
lines. Four absorption lines 'A, B, C, D' revealed intense signals,
whereas the absorption line 'F' is characterized by an narrow width
and a tiny intensity. It turned out that the rotation of the applied
magnetic field $\mathbf{H}$ within the $\mathbf{a,b}$--plane does
not change the position of the absorption lines. The resonance
positions of these absorption lines 'A', $\dots$, and 'F' are
plotted in Fig. \ref{Fig_6} as a function of the magnetic field $H$.
The resonance positions of the narrow line with small intensity,
which is marked with the letter 'F' (the grey squares in Fig.
\ref{Fig_6}), most probably have to be attributed to an electron
paramagnetic resonance (EPR) of impurities in the sample. This
assumption is corroborated by the observation of an decreasing
intensity towards higher temperatures, but temperature independent
resonance positions of these lines 'F'. Therefore, the absorption
lines 'F' will not be discussed in our paper.

\begin{figure}
\includegraphics[width=70mm]{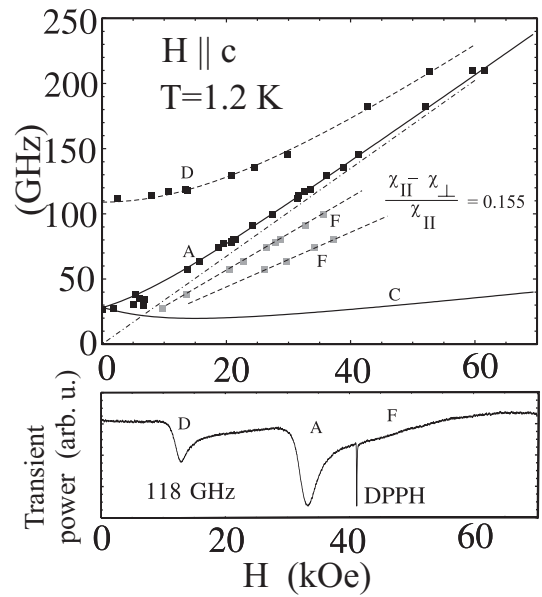}
\caption{Upper panel: AFMR spectra in the 3D ordered phase vs.\
applied magnetic field $H$ ($\mathbf{H}\parallel \mathbf{c}$ at
1.2~K). The capital letters 'A, D, F' mark different branches of the
spectra (see text). Lower panel: one example of the trace of the
transient microwave power vs.\ applied magnetic field $H$ at 118 GHz
($\mathbf{H}\parallel \mathbf{c}$ at 1.2~K). The signal of the
reference compound DPPH is marked.}
 \label{Fig_7}
\end{figure}

Within the field region $22 \lesssim H\lesssim 28$~kOe, which is
marked in light grey in Figs. \ref{Fig_5} and \ref{Fig_6}, a
peculiarity in the AFMR spectra was observed: towards lower fields
$H$, the branch 'A' transforms into the branch 'B' exhibiting a
coexistence of both absorption lines. The absorption line 'A'
appears at slightly higher fields, and the absorption line 'B' at
slightly lower fields with respect to the resonance of the reference
compound diphenyl--picryl--hydrazyl (DPPH, $g=2$), respectively (cf.
Fig. \ref{Fig_5}). This peculiarity for applied magnetic fields
around 25~kOe we associate with a phase transition, most probably of
the spin--flop type at $H_{c1}\approx 25$~kOe. For the orientation
of the applied magnetic field $\mathbf{H}\parallel \mathbf{c}$, two
branches 'A' and 'D' of the AFMR spectra were observed as shown in
figure \ref{Fig_7}. The dashed--dotted lines in Figs. \ref{Fig_6}
and \ref{Fig_7} show the branches of EPR spectra, which were
measured at 4.2 K. The values of the $g$--factors at 4.2 K are in a
good agreement with the values reported in reference
\onlinecite{Krug}. The solid lines in Figs. \ref{Fig_6} and
\ref{Fig_7} show the AFMR spectra which were deduced from a
phenomenological theory for model systems as explained later in the
section \ref{discussion AFMR}. The dashed lines in Figs. \ref{Fig_6}
and \ref{Fig_7} are drawn for simplicity of perception.

\begin{figure}
\includegraphics[width=70mm]{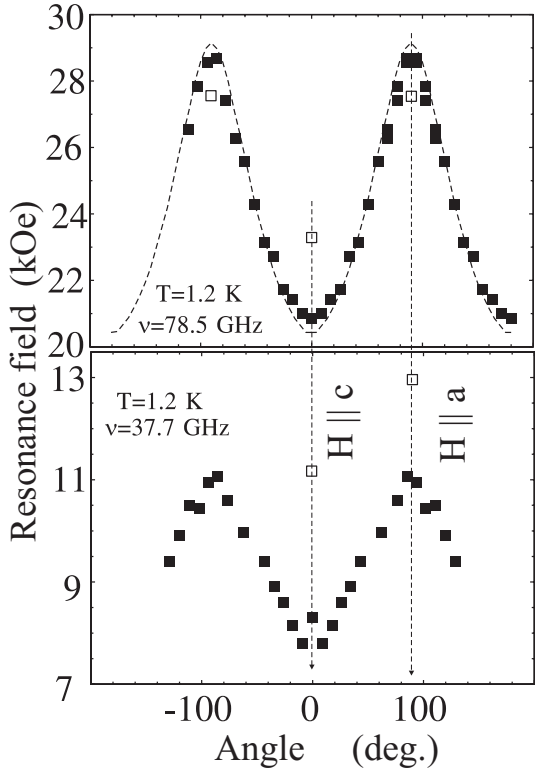}
\caption{Resonance field of the 'A' branch vs.\
$\phi=\angle(\mathbf{H},\mathbf{c})$ at $T=1.2$~K. The field
$\mathbf{H}$ is applied within the $\mathbf{ac}$--plane. The open
squares mark values of EPR fields, measured in the paramagnetic
phase ($T=4.2$~K) for $\mathbf{H}\parallel\mathbf{a}$ and
$\mathbf{H}\parallel\mathbf{c}$, respectively. Upper panel:
$\nu$=78.5 GHz; lower panel: $\nu$=37.7 GHz.}
 \label{Fig_8}
\end{figure}

Figure \ref{Fig_8} shows the dependence of the resonance field of
the 'A'--branch on the angle $\phi$ at the temperature of 1.2~K,
where $\phi$ is the angle between the direction of the applied
magnetic field $\mathbf{H}$ and the $\mathbf{c}$--axis. The field
$\mathbf{H}$ is applied within the $\mathbf{ac}$--plane. The open
squares mark values of EPR fields, measured in the paramagnetic
phase at $T=4.2$~K for $\mathbf{H}\parallel\mathbf{a}$ and
$\mathbf{H}\parallel\mathbf{c}$, respectively. At 78.5 GHz (upper
panel), the resonance field for the 'A'--branch occurs at a field
value which is smaller than the field of the EPR signal, when the
field $\mathbf{H}$ is applied along the $\mathbf{c}$--axis
($\mathbf{H}\parallel\mathbf{c}$). In case of the orientation
$\mathbf{H}\parallel\mathbf{a}$,  the resonance field for the
'A'--branch slightly exceeds the field of the EPR signal. The lower
panel of Fig. \ref{Fig_8} displays the situation at a frequency of
37.7 GHz. Here, the resonances according to this frequency take
place at fields less than the transition field $H_{c1}\thickapprox
25$~kOe. In this case, the resonance fields for all angles $\phi$
are less than the EPR fields. Note that the angular dependences of
the resonance fields predominantly are defined by the angular
dependence of the $g$--factor.

\begin{figure}
\includegraphics[width=70mm]{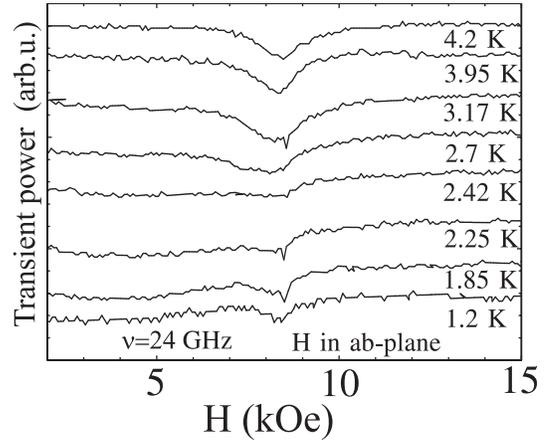}
\caption{Transient microwave power vs.\ applied magnetic field $H$
for different temperatures at 24~GHz. The field $\mathbf{H}$ is
applied within the $\mathbf{ab}$--plane. }
 \label{Fig_9}
\end{figure}

\begin{figure}
\includegraphics[width=70mm]{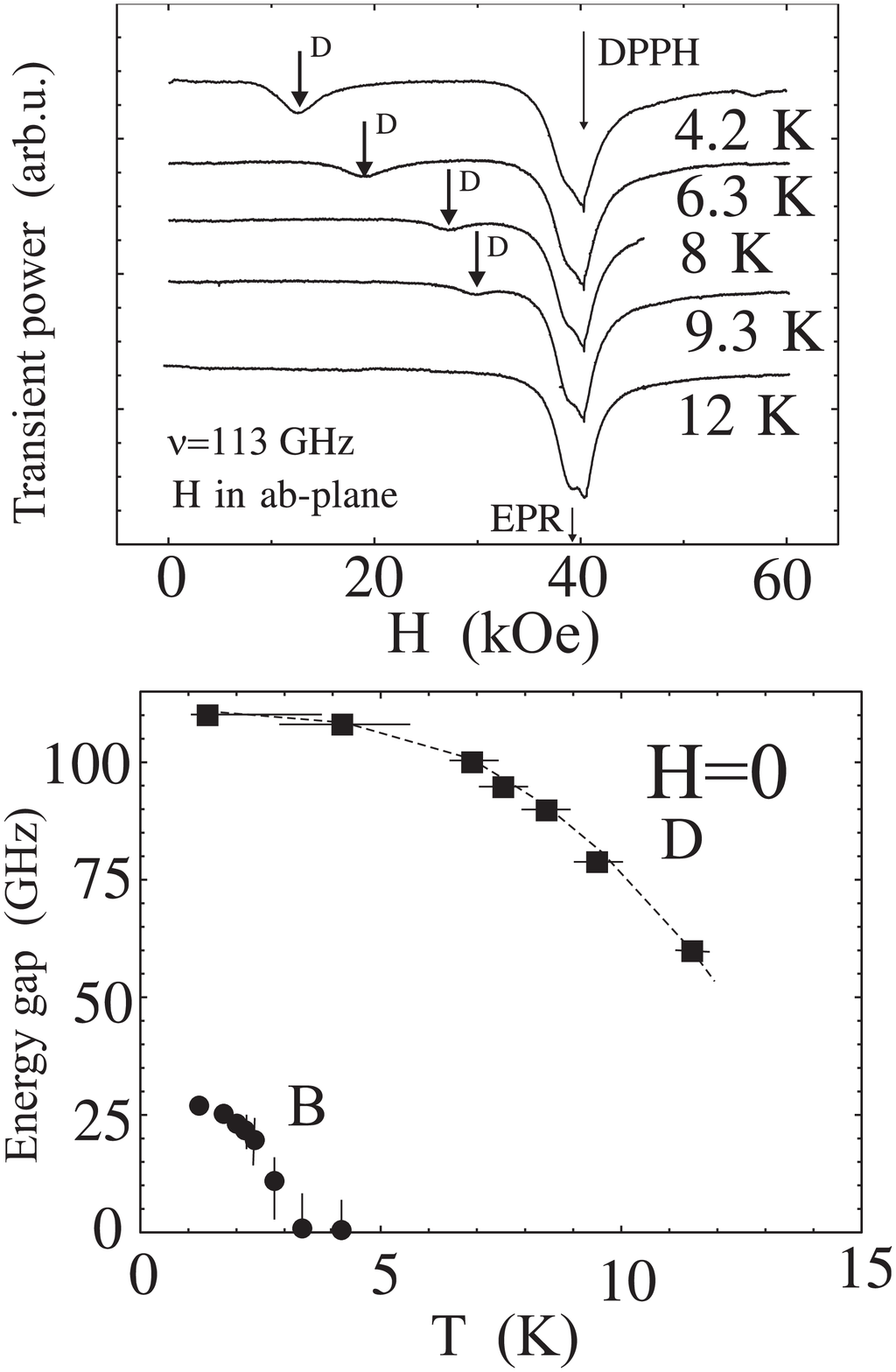}
\caption{Upper panel: transient microwave power vs.\ applied
magnetic field $H$ for different temperatures at 113~GHz
($\mathbf{H} \parallel \mathbf{a}$). Lower panel: temperature
dependence of the AFMR positions of 'B' and 'D' branches at zero
field $H$=0.}
 \label{Fig_10}
\end{figure}

On closer inspection of the experimental AFMR data in Figs.
\ref{Fig_6} and \ref{Fig_7} for applied fields $H\longrightarrow 0$,
there are two branches which exhibit an frequency--axis intercept of
$\nu (H=0)\approx 25$~GHz and $\approx 108$~GHz, respectively. These
two intercepts at 1.2~K deep in the 3D magnetically ordered phase
identify two excitation gaps which turned out to have different
temperature dependences: the low--frequency gap decreases with
increasing temperatures, and the according branch becomes gapless at
a temperature near T$_N$. In Fig. \ref{Fig_9}, traces of the
transient microwave power are displayed as a function of the applied
magnetic field $H$ for different temperatures at $\nu$=24~GHz.
Towards decreasing temperatures, the AFMR line starts to broaden at
a temperature $T \simeq T_N$ and shifts toward lower fields. At 1.2
K, the broad line at zero field $H$=0 shows that the probing
frequency of 24~GHz is almost equal to the gap value at this
temperature. Using equation \ref{eq:2}, analogous spectra obtained
at the probing frequency of $\nu$=38 GHz  allow to obtain the
temperature dependence of the gap value, which is shown in the lower
panel of figure \ref{Fig_10}.
The high--frequency gap, which has to be attributed to the
'D'--branch, remains to be open till much higher temperatures than
$T_N$. The upper panel of Fig. \ref{Fig_10} shows traces of the
transient microwave power as a function of the applied magnetic
field $H$ for different temperatures at $\nu$=113 GHz
($\mathbf{H}\parallel\mathbf{a}$). The AFMR position of the
'D'--branch is well defined till $T\approx 10$~K. The temperature
dependence of this shift of the AFMR position is documented in the
lower panel of figure \ref{Fig_10}. The continuous decrease of the
AFMR frequency towards higher temperatures indicates the closing of
the gap. From the extrapolation of this continuous decrease to zero
frequency we expect the 'D' branch to be gapless at the temperature
of $T\simeq 13$~K.\\

\section{NMR}
\begin{figure}
\includegraphics[width=70mm]{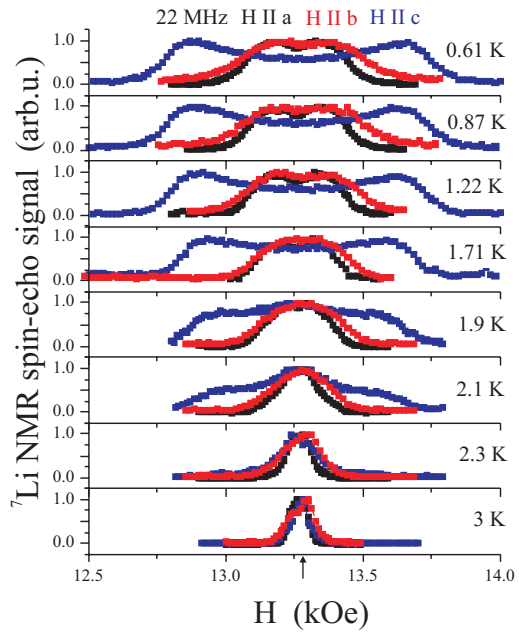}
\caption{Temperature dependence of $^7$Li NMR spectra at 22~MHz. The
applied magnetic field $\mathbf{H}$ was oriented along all
crystallographic axes $\mathbf{a},\mathbf{b},$ and $\mathbf{c}$.}
 \label{Fig_11}
\end{figure}

\begin{figure}
\includegraphics[width=70mm]{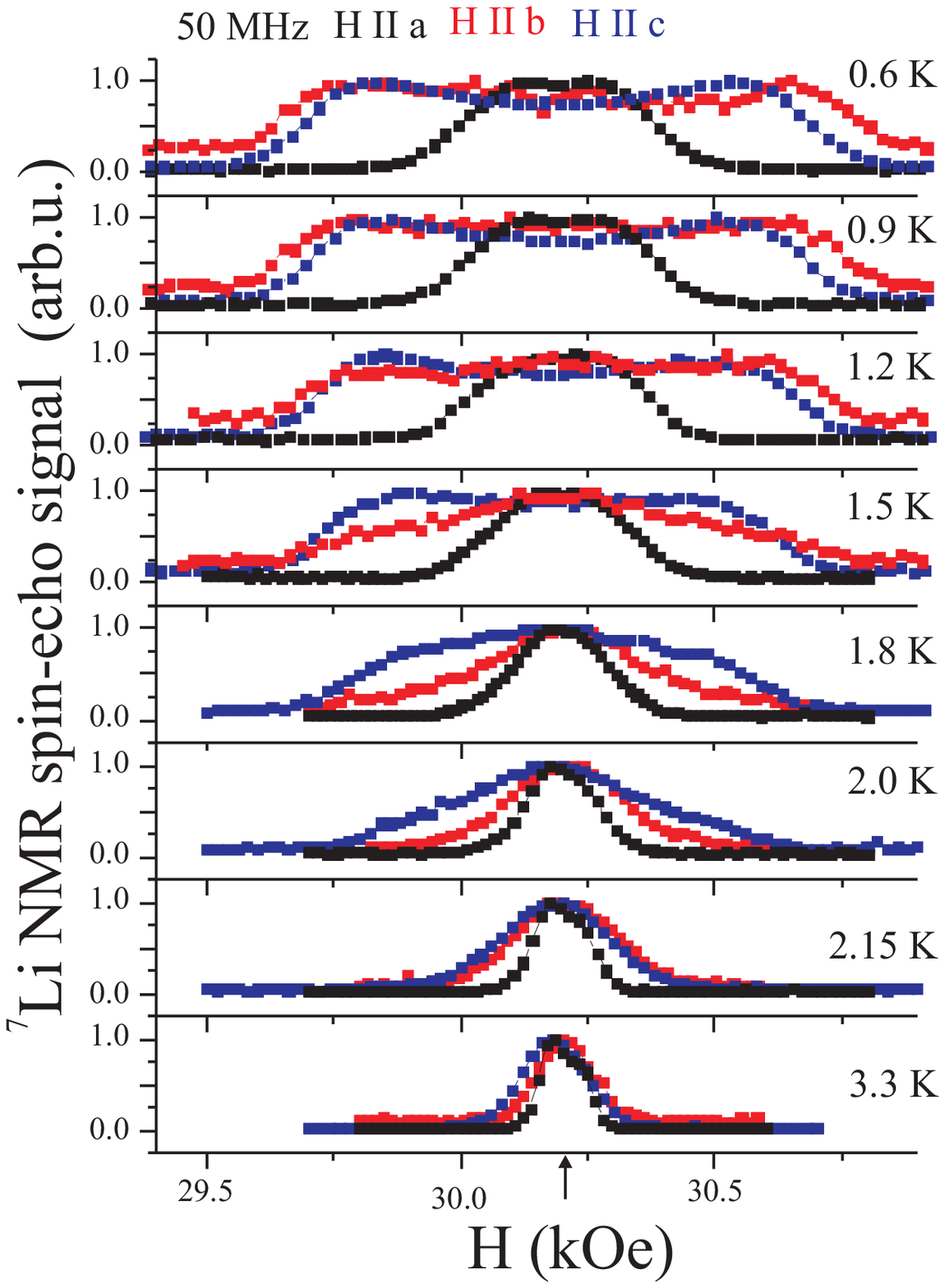}
\caption{Temperature dependence of $^7$Li NMR spectra at 50~MHz. The
applied magnetic field $\mathbf{H}$ was oriented along all
crystallographic axes $\mathbf{a},\mathbf{b},$ and $\mathbf{c}$.}
\label{Fig_12}
\end{figure}

Figures \ref{Fig_11} and \ref{Fig_12} show $^7$Li NMR spectra for
different temperatures obtained at the frequencies of $\nu =22$~MHz
and 50~MHz, respectively. The magnetic field $H$ was applied along
all crystallographic axes ($\mathbf{H}\parallel \mathbf{a}$,
$\mathbf{H}\parallel \mathbf{b}$, and $\mathbf{H}\parallel
\mathbf{c}$). These frequencies correspond to applied magnetic
fields $H$ which are lower than $H_{c1}\approx 25$~kOe in the case
of 22~MHz, and higher than $H_{c1}$ in the case of 50~MHz. For
temperatures within the paramagnetic phase ($T\geq 3$~K), the
spectra for both frequencies are almost unshifted with respect to
the reference field $H_{\rm ref}=\nu/\gamma$ which is indicated by
an arrow in figures \ref{Fig_11} and \ref{Fig_12}. For decreasing
temperatures towards $T_{\rm N}$, the spectra start to broaden. For
temperatures lower than $T_{\rm N}$, the spectra change to a
plateau--like shape with some increase of the intensity at their
borders. Such shape of NMR spectra is characteristic for a spiral
structure of the ordered electron moments, in which the magnetic
ions induce different effective magnetic fields at the probing NMR
nuclei, where the nuclei occupy crystallographically equivalent
sites. From the fact of the temperature independent NMR line shape
for temperatures $0.6 < T< 1.2$~K we conclude that the AFMR
experiments at 1.2~K were performed with nearly saturated magnetic
Cu$^{2+}$ moments.

At 22~MHz for applied fields $H<H_{c1}$ (Fig. \ref{Fig_11}), the NMR
spectrum at 0.6~K by application of the magnetic field along
$\mathbf{c}$ was approximately three times broader than the spectra
observed by orientation of the field along $\mathbf{a}$ and
$\mathbf{b}$. We also measured at the frequency of 14.5~MHz
(corresponding to an applied magnetic field around 8.8~kOe, not
shown) and it turned out that the line shape at 0.6~K for
$\mathbf{H}\parallel\mathbf{b}$ is not changed compared to the
spectrum obtained at 22~MHz (cf. uppermost frame of Fig.
\ref{Fig_11}). At 50~MHz for applied fields exceeding $H_{c1}$ (Fig.
\ref{Fig_12}), the broad characteristic NMR spectra shape was
observed for orientations of the magnetic field $\mathbf{H}$
parallel to $\mathbf{b}$ {\em and} $\mathbf{c}$. For the orientation
$\mathbf{H}\parallel\mathbf{a}$, the NMR spectrum remains to be
about three times narrower.

\begin{figure}
\includegraphics[width=70mm]{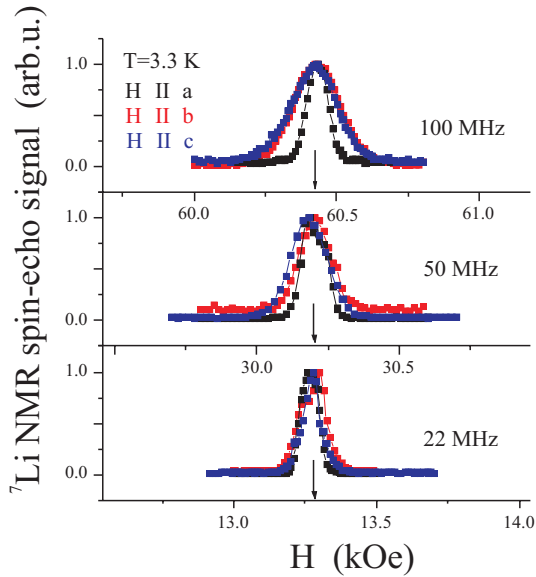}
\caption{Frequency dependence of $^7$Li NMR spectra in the
paramagnetic phase at 3.3~K. The applied magnetic field $\mathbf{H}$
was oriented along all crystallographic axes
$\mathbf{a},\mathbf{b},$ and $\mathbf{c}$.} \label{Fig_13}
\end{figure}

\begin{figure}
\includegraphics[width=70mm]{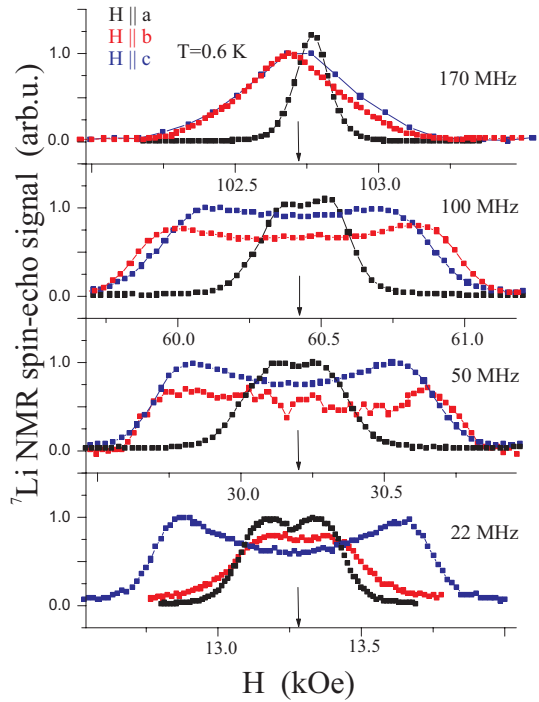}
\caption{Frequency dependence of $^7$Li NMR spectra in the 3D
magnetically ordered phase at 0.6~K. The applied magnetic field
$\mathbf{H}$ was oriented along all crystallographic axes
$\mathbf{a},\mathbf{b},$ and $\mathbf{c}$. } \label{Fig_14}
\end{figure}

Figures \ref{Fig_13} and \ref{Fig_14} give the frequency/field
dependence of $^7$Li NMR spectra in the paramagnetic phase at $T
=3.3 $~K and in the 3D magnetically ordered phase at $T = 0.6$~K,
respectively. In the paramagnetic phase at $T=3.3$~K (cf. Fig.
\ref{Fig_13}), we observed one single, unsplit spectral line for all
frequencies and field orientations and a minor line shift with
respect to the diamagnetic reference field $H_{\rm ref}=\nu/\gamma$
which is indicated by an arrow in figure \ref{Fig_13}. For the field
orientation $\mathbf{H}\parallel\mathbf{a}$, there is no
frequency/field dependence of the $^7$Li NMR line width, whereas for
$\mathbf{H}\parallel\mathbf{b}$ and $\mathbf{H}\parallel\mathbf{c}$
an appreciable increase of the $^7$Li NMR line width became apparent
towards higher frequencies/fields as clearly seen in the uppermost
frame of figure \ref{Fig_13}. In the 3D magnetically ordered phase
at $T=0.6$~K (cf. Fig. \ref{Fig_14}), the line width of the
characteristic spectra shape for $\mathbf{H}\parallel\mathbf{c}$
turned out to be frequency/field independent up to the frequency of
100~MHz and/or field of 60.5~kOe, respectively. However, the line
intensity of the central part of these spectra increases with
increasing frequency/field. At 100~MHz/60.5~kOe, a plateau--like
shape evolved from the double-horn shape at 22~MHz/13.25~kOe. As the
frequency/field is further increased to 170~MHz/102.75~kOe, the
shape of the $^7$Li NMR spectra changes to one single, unsplit
spectral line (uppermost frame of Fig. \ref{Fig_14}). As high--field
magnetization measurements revealed a spin reorientation taking
place at an applied magnetic field of $H_{c2}\approx 75$~kOe (Ref.
\onlinecite{Bank}), we interpolate for the change to one single,
unsplit spectral line to take place at $H_{c2}\approx 75$~kOe.

\begin{figure}
\includegraphics[width=70mm]{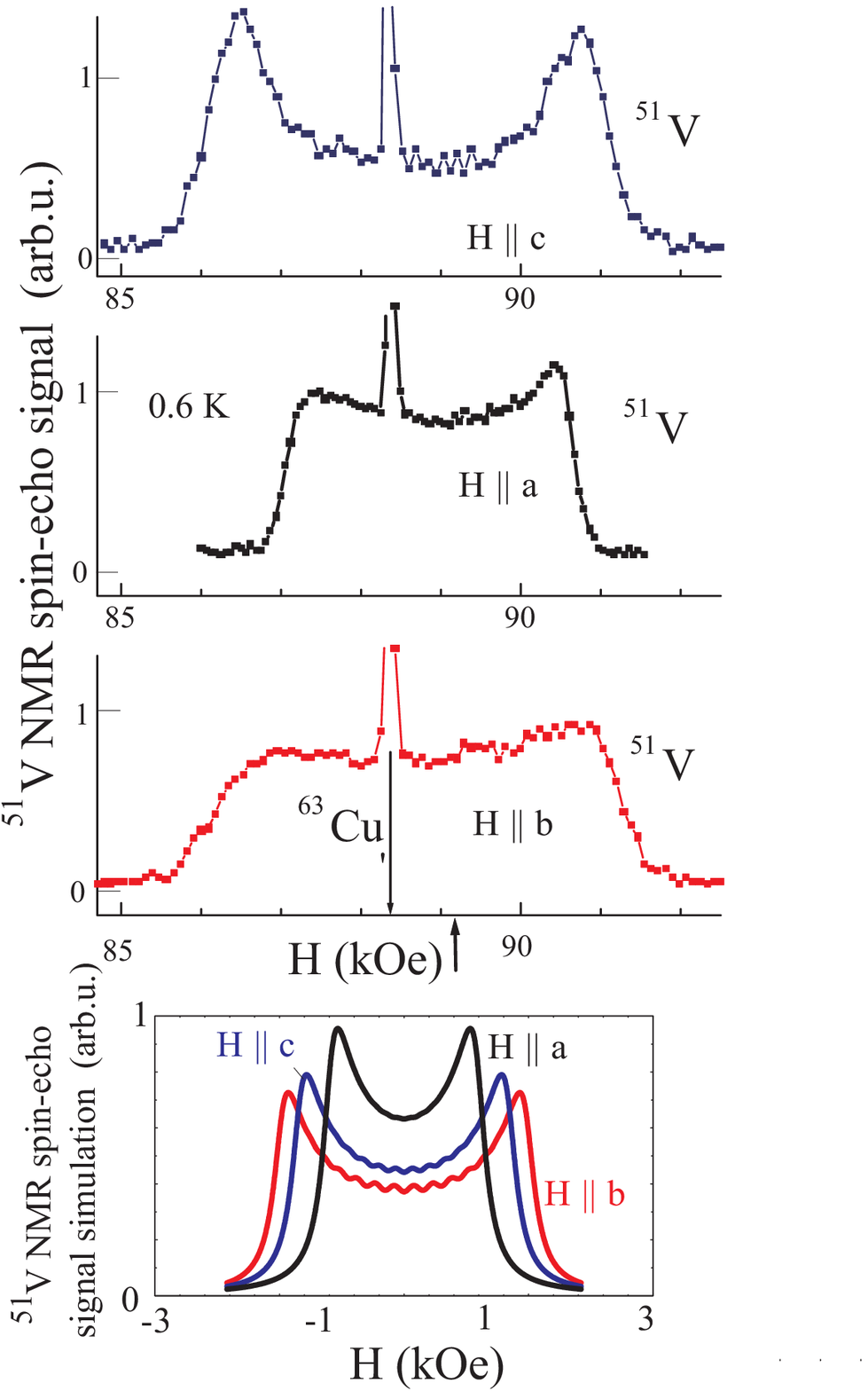}
\caption{Upper panel: orientation dependence of $^{51}$V NMR spectra
in the 3D magnetically ordered phase at 0.6~K and 100~MHz. The
applied magnetic field $\mathbf{H}$ was oriented along all
crystallographic axes $\mathbf{a},\mathbf{b},$ and $\mathbf{c}$.
Lower panel: computed NMR spectra.} \label{Fig_15}.
\end{figure}

Figure \ref{Fig_15} shows the orientation dependence of $^{51}$V NMR
spectra in the 3D magnetically ordered phase at 0.6~K for
$\nu=100$~MHz. The intense narrow lines at around 88.4~kOe are due
to the $^{63}$Cu NMR signal from the copper wire of the resonant
circuit. For all three crystallographic axes
$\mathbf{a},\mathbf{b},$ and $\mathbf{c}$, the line splitting of the
$^{51}$V NMR spectra amounts to five times the splitting of the
$^{7}$Li NMR spectra. The characteristic NMR line shape of $^{51}$V
for $\mathbf{H}\parallel\mathbf{c}$ has also been observed in a
recent NMR study at 120~kOe (Ref. \onlinecite{Smit}). Due to a much
stronger hyperfine coupling of the $^{51}$V nuclei,\cite{Kegl} the
effective magnetic fields induced by the Cu$^{2+}$ moments are
strongly enhanced compared to the effective local fields induced at
the $^7$Li nuclei. We also tried to detect $^{51}$V NMR spin--echo
signals in the low--field range for $H<H_{c1}$, but for all three
crystallographic axes $\mathbf{a},\mathbf{b},$ and $\mathbf{c}$ our
attempts were unsuccessful. We attribute the lack of a vanadium
signal to very short spin--spin relaxation times in this field
range.

\section{DISCUSSION OF NMR EXPERIMENTS}
\label{discussion NMR}

The magnetic structure of LiCuVO$_4$ is defined by strongly bonded
chains of magnetic Cu$^{2+}$ ions, directed along the
crystallographic $\mathbf{b}$--axis. The intra-- and interchain
exchange integrals are defined in Ref. \onlinecite{Ende}: the
intrachain exchange integrals $J_1$ and $J_2$ between nearest and
next--nearest moments of copper ions in the chain are -1.6 and
5.6~meV, respectively (cf. Fig. \ref{Fig_1}). The interchain
exchange integrals $J_3$, $J_4$, $J_5$, and $J_6$ between chains
given in Fig. \ref{Fig_2} amount to 0.01, -0.37, -0.014 and
0.08~meV, respectively.

The interchain interactions between the chains along the direction
of the $\mathbf{a}$--axis are defined by the exchange integrals
$J_3$ and $J_4$ yielding a co--ordered arrangement of magnetic
moments of neighboring copper ions in this direction. This
interaction is much stronger than the interaction along
$\mathbf{ac}$--diagonal ($J_5$, $J_6$ in Fig. \ref{Fig_2}), which
define the interchain interaction of magnetic moments in neighboring
$\mathbf{ab}$--planes. The simple evaluation of the exchange
interactions with these values of exchange integrals give the
following hierarchy of energy of interactions recomputed on one
magnetic ion: the ferromagnetic interaction between the chains
within $\mathbf{ab}$--planes is about one order of magnitude less
than the intrachain interactions and around ten times larger than
the antiferromagnetic interaction between neighboring
$\mathbf{ab}$-planes. The magnetic moments of Cu$^{2+}$ ions in
neighboring $\mathbf{ab}$--planes are oriented
antiparallel.\cite{Gibs}

In the present paper we study NMR on $^7$Li and $^{51}$V nuclei. In
Fig. \ref{Fig_2}, the positions of one Li$^+$ ion and one V$^{5+}$
ion are shown exemplarily, and it turns out that the lattice
positions of these ions are essentially different: the position of
the Li$^+$ ions contains the operation of inversion, but the
position of V$^{5+}$ does not. The $^7$Li nuclei reside on positions
between neighboring $\mathbf{ab}$--planes. Every lithium ion has
four nearest copper ions, which are equally distanced from it. Each
vanadium ion resides between six neighboring copper ions, at that,
four of them belong to one $\mathbf{ab}$-plane, and two, more
distant, to another $\mathbf{ab}$-plane.

The effective local fields at the probing nuclei of the nonmagnetic
lithium and vanadium ions are composed of the long--range dipole
fields of surrounding magnetic moments and of the so called
"contact"\ hyperfine fields due to the nearest magnetic ions. The
contact fields are expected to be proportional to the spin values of
the neighboring copper ions. The contact fields at the $^7$Li nuclei
must be self--compensated in the case of the antiparallel magnetic
ordering of the nearest magnetic moments of neighboring
$\mathbf{ab}$--planes. At the $^{51}$V nuclei, the contact fields
will be predominantly given by the spins of the four nearest
magnetic ions of the nearest neighboring $\mathbf{ab}$--plane (cf.
Fig. \ref{Fig_2}). The value of the long--range dipole field  was
computed for the positions of $^7$Li and $^{51}$V nuclei by
assumption of the spiral magnetic structure proposed in reference
\onlinecite{Gibs}. For this structure, the magnetic moments of the
copper ions can be calculated:
\begin{eqnarray}\label{eq:1}
\mathbf{\mu}=\mu_{\rm ord} \cdot \mathbf{l_1} (-1)^{2z/c} \cdot
\cos(k_{ic}\cdot y)+
\\ \nonumber +\mu_{\rm ord} \cdot \mathbf{l_2} (-1)^{2z/c} \cdot \sin(k_{ic}\cdot
y),
\end{eqnarray}
where the coordinates $x,y,$ and $z$ are given in the basis of the
crystal axes $\mathbf{a}$,$\mathbf{b}$, and $\mathbf{c}$. The
vectors $\mathbf{l_1}$ and $\mathbf{l_2}$ are orthogonal unit
vectors within the $\mathbf{ab}$--plane. Consequently, the vector of
the exchange structure $\mathbf{n}$=$\mathbf{l_1} \times
\mathbf{l_2} $ is parallel to the $\mathbf{c}$--axis. The values of
the lattice parameters $a,b$ and $c$, and the coordinates of lithium
and vanadium ions where taken from reference \onlinecite{Kann}. For
the value of the ordered magnetic moment $\mu_{\rm ord}$ of copper
ions we took 0.31 $\mu_B$ as it was obtained in reference
\onlinecite{Gibs}.\\

\begin{figure}
\includegraphics [width=70mm]{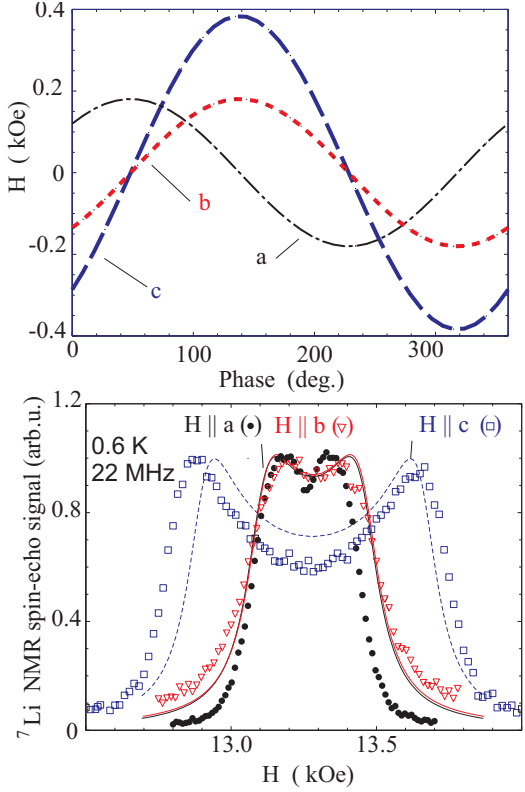}
\caption{Upper panel: computed projections of dipole fields along
the $\mathbf{a}$, $\mathbf{b}$, and $\mathbf{c}$ axes as a function
of the phase of the spiral magnetic structure
$\mathbf{n}\parallel\mathbf{c}$ suggested in Ref. \onlinecite{Gibs}.
Lower panel: experimental $^7$Li NMR spectra measured at $T=0.6$~K
for $H<H_{c1}$ (symbols) and computed NMR spectra (dashed and solid
lines).} \label{Fig_16}
\end{figure}

\noindent $H<H_{c1}$: \\
The dipole field at the $^7$Li nuclei depends on their positions
within the incommensurate spiral magnetic structure. In the upper
panel of Fig. \ref{Fig_16}, the projections of dipole fields on the
$\mathbf{a}$--, $\mathbf{b}$--, and $\mathbf{c}$--axes are plotted
as a function of the phase of the spiral magnetic structure.  From
these fields it is simple to simulate the expected NMR spectra for
all three directions of the applied magnetic field $H$ with respect
to the crystallographic axes. The lower panel of Fig. \ref{Fig_16}
shows computed spectra (solid lines) and experimental data (symbols)
for the frequency of 22~MHz. This computation was performed with a
Lorentzian shape of the NMR spectrum which is produced by the group
of $^7$Li nuclei experiencing the same value of the effective local
magnetic field. The line width of this Lorentzian--shaped spectrum
was assumed to be equal to 75 Oe. This value of the line width is
consistent with the line width in the paramagnetic phase of the
low--field range (Fig. \ref{Fig_11}). The computed NMR spectra are
in a very good agreement with the experimental spectra. That means
that the effective local fields at the $^7$Li nuclei are
predominantly defined by long--range dipole fields, and we should be
able to specify the magnetic structures of LiCuVO$_4$ for elevated
applied magnetic fields $H>H_{c1}$.\\

\begin{figure}
\includegraphics [width=70mm]{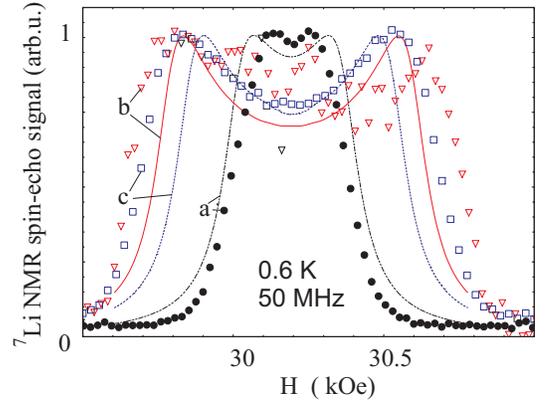}
\caption{Experimental $^7$Li NMR spectra measured at $T=0.6$~K and
$H>H_{c1}$ (symbols) and computed NMR spectra (dashed and solid
lines).} \label{Fig_17}.
\end{figure}

\noindent $H_{c1}<H<H_{c2}$: \\
The most natural magnetic transition taking place with increasing
applied magnetic fields $H$ is the spin--flop transition, i.e. the
rotation of the plane of the spiral spin structure perpendicular to
the direction of the applied magnetic field $H$. Therefore, for
$\mathbf{H}\parallel\mathbf{c}$ there will be no spin reorientation
expected. Indeed, the experiment exhibits nearly the same pattern of
the NMR spectra for $\mathbf{H}\parallel\mathbf{c}$ at all
frequencies 22~MHz, 50~MHz, and 100~MHz, corresponding to applied
magnetic fields around $H_{c1}\approx 25$~kOe (cf. Fig.
\ref{Fig_14}). Applying a magnetic field $H>H_{c1}$ parallel to the
$\mathbf{a}$-- and $\mathbf{b}$--axes may result in a rotation of
the magnetic structure and the line shapes concomitantly can change.
Figure \ref{Fig_17} shows the results of the computation of the NMR
spectra for these structures together with the experimental NMR
spectra measured at the frequency of 50~MHz. The computation and
experiment show that the spin reorientation drastically broadens the
NMR spectra for $\mathbf{H}\parallel \mathbf{b}$, and nearly no
change is observed for $\mathbf{H}\parallel \mathbf{a}$. This
magnetic structure is maintained up to applied magnetic fields
$H\approx 60.5$~kOe $<H_{c2}$, as it is seen from the NMR spectra
pattern at 100~MHz/60.5 kOe in figure \ref{Fig_14}. Thus for
$H_{c1}<H<H_{c2}$, the magnetic structure exhibits the spiral plane
to be perpendicular to the applied magnetic field $\mathbf{H}
\parallel \mathbf{n}$ (cf. Eq. \ref{eq:1}) for $\mathbf{H}
\parallel \mathbf{b}$ and $\mathbf{H}\parallel \mathbf{c}$, respectively. The agreement
of the experimental and computed spectra only was possible to obtain
in the case of an antiferromagnetic alternation of the magnetic
moments of neighboring $\mathbf{ab}$--planes (cf. Fig.
\ref{Fig_2}).\\

\noindent $H>H_{c2}$: \\
A more complicate situation occurs at high fields $H> H_{c2}$: the
NMR spectra of $^7$Li nuclei in this field range consist of one
single line for all field directions (cf. the uppermost panel in
Fig. \ref{Fig_14}), indicating a second spin reorientation process
to take place. For the high field range $H> H_{c2}$ it was possible
to measure NMR spectra of the $^{51}$V nuclei, too. To understand
the nature of the effective local field at the  $^{51}$V nuclei we
computed the long--range dipole field at the vanadium site for the
paramagnetic phase. The computed dipole field is -0.44 kOe/$\mu_B$
for the direction of the applied field $\mathbf{H}$ parallel to the
$\mathbf{b}$-axis. But the value of the hyperfine coupling constant
${}^{51}$A$_{\parallel b}$=4.95 kOe/$\mu_{B}$ is about ten times
larger as we obtained in reference \onlinecite{Kegl}. Thus we can
conclude that the main part of the effective local field at the
$^{51}$V nuclei is defined by contact fields. The value of the
effective contact field induced by copper ions at vanadium nuclei
can be evaluated as
5.4($\overline{\mu_1}$+$\overline{\mu_2}$+$\overline{\mu_3}$+$\overline{\mu_4}$)/4~kOe/$\mu_B$.
Here we take into account only the four nearest magnetic copper ions
from the nearest neighboring $\mathbf{ab}$--plane (Fig.
\ref{Fig_2}).

The characteristic broad NMR spectra of $^{51}$V nuclei (Fig.
\ref{Fig_15}) in the 3D magnetically ordered phase for $H>H_{c2}$
indicate a modulation of the projection of the Cu$^{2+}$ moments on
the direction of the applied field $\mathbf{H}$. A more simple way
to describe these modulation is to suppose that at $H_{c2}$ the
spiral planar magnetic structure orients to be parallel to the
applied field $H$ (i.e., $\mathbf{n} \perp \mathbf{H}$). We computed
the value of the effective local fields at the $^{51}$V nuclei using
the same parameters k$_{ic}$ and $\mu_{\rm ord}$ of the spiral
magnetic structure at zero applied field $H$=0. The computed NMR
spectra with these parameters are shown in the lowest panel of
figure \ref{Fig_15}. Again, this computation was performed with a
Lorentzian shape of the NMR spectrum which is produced by the group
of $^{51}$V nuclei experiencing the same value of the effective
local magnetic field. The line width of this Lorentzian--shaped
spectrum was assumed to be equal to 150 Oe. Note that the computed
NMR spectra better fit, if the value of $\mu_{\rm ord}$ is increased
and/or the value of k$_{ic}$ is decreased.

\begin{figure}
\includegraphics [width=75mm]{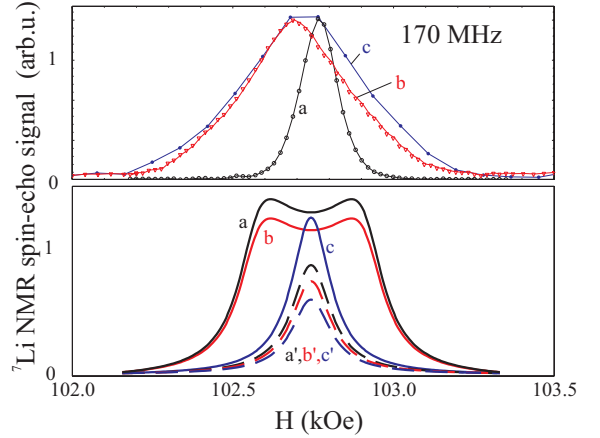}
\caption{Upper panel: experimental $^7$Li NMR spectra measured at
$T=0.6$~K and $H>H_{c2}$. The applied magnetic field $\mathbf{H}$
was oriented along all crystallographic axes
$\mathbf{a},\mathbf{b},$ and $\mathbf{c}$. Lower panel: computed NMR
spectra for spiral-- (solid lines, $\mathbf{a}$, $\mathbf{b}$, and
$\mathbf{c}$) and spin--modulated structure (dashed lines,
$\mathbf{a'}$, $\mathbf{b'}$, and $\mathbf{c'}$), respectively.}
\label{Fig_18}.
\end{figure}

Changing again to the $^7$Li NMR spectra for $H>H_{c2}$ (cf. upper
panel in Fig. \ref{Fig_18}), the unsplit line shapes of the NMR
spectra suggest the following magnetic structures which are deduced
from computations of dipole fields for different planar spiral
structures: for $\mathbf{H}\parallel\mathbf{c}$, the best fit is
obtained for $\mathbf{n}\parallel \mathbf{b}$. For
$\mathbf{H}\parallel(\mathbf{a,b})$, the plane of the spiral
structure lies within the $\mathbf{ab}$--plane (i.e.,
$\mathbf{n}\parallel \mathbf{c}$). In the lower panel of Fig.
\ref{Fig_18} the results of the computed NMR spectra are shown for
these spiral structures. For field orientations parallel to the
$\mathbf{a}$-- and $\mathbf{b}$--axes, the computed spectra exhibit
some splitting of the spectra shape contrary to the experiment. Due
to this lack of any splitting of the experimental spectra for
$H>H_{c2}$ we conclude that {\em not} the spiral--, but a
spin--modulated structure is realized, i.e. the spin projections
perpendicular to the direction of the applied field $H$ are
disordered. The NMR spectra computed for this spin--modulated
structure are shown in the same figure (dashed lines in the lower
panel of Fig. \ref{Fig_18}). All spectra for this structure are
unsplit. This computation was performed with a line width of 150~Oe
of the Lorentzian--shaped NMR spectrum which is produced by the
group of $^7$Li nuclei experiencing the same value of the effective
local magnetic field. In order to achieve a better fit for field
directions $\mathbf{H}\parallel (\mathbf{b,c})$, this value of the
line width has to be increased by a factor of 2.8. At first glance
this increase seems to be arbitrary, but on closer inspection of the
experimental spectra in the paramagnetic phase (uppermost panel of
Fig. \ref{Fig_13}),  the spectral line widths for field directions
$\mathbf{H}\parallel (\mathbf{b,c})$ are also significantly
increased compared to $\mathbf{H}\parallel\mathbf{a}$.

\section{DISCUSSION OF AFMR EXPERIMENTS}
\label{discussion AFMR}

For low applied magnetic fields $H<H_{c1}$, the authors of Ref.
\onlinecite{Gibs} propose the planar spiral structure for the
magnetic structure of LiCuVO$_4$ in the 3D magnetically ordered
phase. This structure was found to be defined by strong exchange
interactions within the magnetic chains.\cite{Ende} Our results of
the NMR and susceptibility $\chi$ investigations can be explained on
the same footing. The results of EPR (Ref. \onlinecite{Krug}) and
AFMR (this work) hint towards an uniaxial character of the crystal
anisotropy in LiCuVO$_4$. These observations motivate to discuss the
low--field AFMR data in the assumptions of an exchange--rigid planar
uniaxial magnetic structure with an uniaxial crystal anisotropy,
which is directed along the $\mathbf{c}$-axis. Thus, we will suppose
that the susceptibility of the exchange structure will be defined by
two values $\chi_{\parallel}$ and $\chi_{\perp}$, i.e. along and
perpendicular to the vector of the exchange structure $\mathbf{n}$
(cf. Eq. \ref{eq:1}), respectively. The anisotropy energy within
this model can be written in a form $\beta$n$_{z}^2$/2 ($\beta<0$).
The NMR and AFMR data reveal a magnetic transition at $H_{c1}$, if
the magnetic field $H$ is applied parallel to the
$\mathbf{ab}$--plane. This implies that $\chi_{\parallel}$ is larger
than $\chi_{\perp}$, and the transition field can be described as
$H_{c1}^2=\beta/(\chi_\perp-\chi_{\parallel}$).

In Ref. \onlinecite{Zali}, the low--frequency branches of the AFMR
spectra for noncollinear complanar structures with an uniaxial
anisotropy were obtained in the framework of a phenomenological
theory according to reference \onlinecite{Andr}. The spectra of
acoustic modes consist of three branches, which correspond to three
rotational degrees of freedom. In the case of an uniaxial exchange
structure, one branch out of these three branches must be zero
($\nu$=0), which is attributed to the rotational degree of freedom
of the exchange structure around the $\mathbf{n}$--axis. The
frequencies of the two other branches can be obtained from the
square equation:
\begin{eqnarray}\label{eq:2}
(\nu/\gamma)^4- (\nu/\gamma)^2 \{H^2+ \eta^2H^2
\cos^2(\theta-\phi)\\ \nonumber +\eta H_{c1}^2 (3 \cos^2\theta-1) \}
+\eta\{-H_{c1}^2\cos^2\theta \\ \nonumber- H^2\cos^2 (\theta-\phi)\}
\{-\eta H_{c1}^2 \cos2\theta \\ \nonumber -\eta H^2\cos^2
(\theta-\phi) -H^2 \sin^2(\theta-\phi)\}=0,
\end{eqnarray}
where $\gamma = g \mu_B/2 \pi \hbar$, $\eta$=
($\chi_{\parallel}$-$\chi_{\perp}$)/$\chi_{\perp}$ and
$\phi=\angle(\mathbf{H}, \mathbf{c})$ is the angle between the
crystallographic $\mathbf{c}$--axis and the direction of the applied
magnetic field $\mathbf{H}$. The angle $\theta=\angle(\mathbf{n},
\mathbf{c})$ is defined by:
\begin{eqnarray}\nonumber
\tan(2\theta)=\frac{H^2\sin2\phi}{H^2\cos(2\phi)+H_{c1}^2}
\end{eqnarray}
If the applied magnetic field is directed along the
$\mathbf{c}$--axis ($\mathbf{H}\parallel\mathbf{c}$), both angles
are zero ($\theta$=0, $\phi$=0). Whereas,
\begin{equation}
\begin{array}{rl}\nonumber
{\rm for}\;\; \mathbf{H}\perp\mathbf{c}, H<H_{c1}: & \theta=0,
\phi=\pi/2,\\
{\rm for}\;\; \mathbf{H}\perp\mathbf{c}, H>H_{c1}: & \theta=\pi/2,
\phi=\pi/2.
\end{array}
\end{equation}
At zero applied field $H$=0, the frequency of both oscillations must
be $\nu_{1,2}$=$\gamma H_{c1}\sqrt{\eta}$. Taking the frequency of
27$\pm$2~GHz of the AFMR at zero applied field and the field
$H_{c1}$ of the spin--flop transition to be $H_{c1}$=25$\pm$3~kOe,
we obtain the parameter of the anisotropy of the exchange
susceptibility $\eta$=0.155$\pm$0.05, and the value of the
anisotropy constant $\beta =-0.01$~K, recomputed on one copper ion.
The results of the computation of the 'A', 'B', and 'C' branches of
the AFMR spectra for $\eta$=0.155 are shown as solid lines in Figs.
\ref{Fig_6} and \ref{Fig_7}. The value of the anisotropy of the
exchange parameter $\eta$ is in satisfactory agreement with the
value obtained from susceptibility measurements at low fields
(Fig.\ref{Fig_4}). Note, the theory of AFMR spectra discussed above
is developed in an exchange approximation with isotropic $\gamma$.
For the computed spectra shown in Figs. \ref{Fig_6} and \ref{Fig_7}
we used the values of $\gamma_\parallel = 3.25$~GHz/kOe and
$\gamma_\perp =  2.85$~GHz/kOe, which were measured in the
paramagnetic phase. We can conclude that this model describes the
AFMR spectra within the low--field range ($H\ll H_{c2}$) very well.

The magnetic phase $H$--$T$ diagram of LiCuVO$_4$ is rich not only
in field cut, but in temperature also. It seems that the
establishment of the 3D magnetic order appears through some
intermediate phase transition. This we conclude from the fact that
the AFMR branch 'D' in Figs. \ref{Fig_6} and \ref{Fig_7} starts to
be gapped at a temperature much higher than $T_N$ ($\simeq$13 K)
(Fig. \ref{Fig_10}). The fingerprint of this transition around
$T=13$~K additionally has recently been detected in measurements of
the temperature dependence of the heat capacity,\cite{Bank2} where a
peak of $C(T)$ centered at $T=12$~K has been reported.

\section{CONCLUSION}
\label{conclusion}

The NMR and AFMR results can be self--consistently described by the
following phase diagram:
\begin{equation}
\begin{array}{rl}\nonumber
H<H_{c1}: & \mathbf{n} \parallel \mathbf{c}\\
H_{c1}<H<H_{c2}: & \mathbf{n} \parallel \mathbf{H}\\
H_{c2}<H: & \mathbf{l_1}\parallel\mathbf{H} \;\; {\rm and}\;\;
\mathbf{l_2}=0,
\end{array}
\end{equation}
where the magnetic structures are described in terms of equation
\ref{eq:1}. It is important to note that for $H> H_{c2}$ from our
experiments we can conclude surely only about the presence of a
modulation of projections of copper moments along the direction of
the applied magnetic field $H$ with an amplitude of $\approx
0.4\pm$0.1 $\mu_B$. The best fit of  $^7$Li NMR spectra is obtained
based on a spin--modulated structure, i.e. the spin components of
Cu$^{2+}$ perpendicular to the direction of the applied magnetic
field $H$ are disordered.

The low--field phases for $H<H_{c2}$ observed in LiCuVO$_4$ are
natural from the point of view of mean--field theory. The
susceptibility of the exchange structure in the direction
perpendicular to the $\mathbf{a,b}$--plane is larger than the
susceptibility within this plane. The competition of the Zeeman
energy and the energy of the crystal--field anisotropy explains the
low--field transition observed at $H_{c1}$. The second transition at
the field of $H_{c2}$ is unusual. To explain this transition we have
to assume that with an increase of the applied magnetic field that
structure starts to be preferable which exhibits the spin modulation
parallel to the direction of the applied magnetic field, rather than
the structure with the spin plane being perpendicular to the field
direction. Probably, it is necessary to take into account the
thermal and quantum fluctuations in order to explain the situation.
For example, for the planar spiral structure on the triangular
lattice it was found that some energy gain occurs for the
orientation of spin plane to be parallel to the direction of the
applied magnetic field.\cite{Chub}

\begin{acknowledgments}
We are grateful to A. Pimenova and D. Vieweg for susceptibility
measurements. We thank V.N. Glazkov, V.I. Marchenko, A.I. Smirnov,
S.S. Sosin, G.B. Teitel\'baum, and A. Loidl for useful discussions.
This work is supported by the Grants 04-02-17294, 06-02-16509
Russian Foundation for Basic Research, Russian President Program of
Scientific Schools, contract BMBF via VDI/EKM 13N6917, and program
of the German Research Society within the Sonderforschungsbereich
484 (Augsburg).
\end{acknowledgments}

\end{document}